\documentstyle[sprocl,psfig,epsfig]{article}

\def\Journal#1#2#3#4{{#1} {\bf #2}, #3 (#4)}

\def\NPB{{\em Nucl. Phys.} B}
\def\PLB{{\em Phys. Lett.}  B}

\def\PRD{{\em Phys. Rev.} D}

\def\ra{\rightarrow}

\def\be{\begin{equation}}
\def\ee{\end{equation}}
\def\ba{\begin{array}}
\def\ea{\end{array}}
\def\bea{\begin{eqnarray}}
\def\eea{\end{eqnarray}}
\def\vev#1{\left\langle #1\right\rangle}
\def\21{$SU(2) \otimes U(1)$}
\newcommand{\ptmis}{{ {\rm p} \hspace{-0.53 em} \raisebox{-0.27 ex} {/}_T }}
\newcommand {\ignore}[1]{}

\begin{document}

\title{SPONTANEOUS BREAKING OF R-PARITY\footnote{
Talk given at {\it International Workshop on 
 Physics Beyond The Standard Model:
 from Theory to Experiment}, Val\`encia, Spain, October 1997.}
}

\author{J. C. Rom\~ao}

\address{Instituto Superior T\'ecnico, Departamento de F\'{\i}sica\\
A. Rovisco Pais 1, 1096 Lisboa Codex, Portugal}

\maketitle\abstracts{
If supersymmetry  is realized with spontaneous breaking of R-parity, 
there will be important consequences in several
different areas which can be tested through different types
of experiments. In this talk we review the phenomenological
implications of these theories, with special emphasis on new signals
at the present and future accelerators.}

\section{Introduction}

So far most attention to the study of supersymmetric phenomenology has
been made in the framework of the Minimal Supersymmetric Standard
Model (MSSM)~\cite{mssm} with conserved R-parity~\cite{RP}.  R-parity
is a discrete symmetry assigned as $R_p=(-1)^{(3B+L+2S)}$, where L is
the lepton number, B is the baryon number and S is the spin of the
state. If R-parity is conserved all supersymmetric particles must
always be pair-produced, while the lightest of them must be stable.
Whether or not supersymmetry is realized with a conserved R-parity is
an open dynamical question, sensitive to physics at a more fundamental
scale.

The study of alternative supersymmetric scenarios where the effective
low energy theory violates R-parity~\cite{HallSuzuki} explicitly
has received recently a lot of attention~\cite{beyond,RPothers}. 
Although highly constrained by proton stability, their systematic 
study at a phenomenological level is hardly possible, due to the 
large number of parameters characterizing these models, in addition 
to those of the MSSM.

As other fundamental symmetries, it could well be that R-parity is a
symmetry at the Lagrangian level but is broken by the ground state.
Such scenarios~\cite{Majoron,model} 
provide a very {\sl systematic} way to include R parity
violating effects, automatically consistent with low energy {\sl
baryon number conservation}. 

In this review talk we will present a viable model for spontaneous
breaking of R-parity and illustrate its phenomenological consequences.

\section{A Viable Model for Spontaneous R-Parity Breaking}

\subsection{The Model}

In the original proposal~\cite{Majoron} the content was just the 
MSSM and the breaking was induced by the $\tilde{\nu}_{\tau}$
acquiring a {\it vev}, $\vev{\tilde{\nu}_{\tau}} = v_L$. As a consequence
the Majoron (J) coupled to the $Z^0$ with gauge strength and the decay
$Z^0 \ra \rho_L J$ contributed with the equivalent of $1/2$ a neutrino
to the invisible $Z^0$ width. As this was ruled out but the LEP
results, a possible way out was proposed. The idea~\cite{model} 
was to enlarge the model and make $J$ mostly out of {\it
isosinglets}. The model is defined by the {\it Superpotential}

\bea
W&=&h_u u^c Q H_u + h_d d^c Q H_d + h_e e^c L H_d \cr
&&+(h_0 H_u H_d - \epsilon^2 ) \Phi \cr
&&+ h_{\nu} \nu^c L H_u + h \Phi \nu^c S 
\label{eq:super}
\eea
where the lepton number assignments are given in Table~\ref{table1}.
\begin{table}[t]
\begin{center}
\begin{tabular}{|l|c|c|c|c|c|} \hline
Field & $L$ & $e^c$ & $\nu^c$ & $S$ & others  \\ \cline{1-6}
Lepton \# & $1$ & $-1$ & $-1$& $1$ & $0$ \\ \hline
\end{tabular}
\caption{Lepton number assignments for the superfields in
Eq.~\ref{eq:super}}
\label{table1}
\end{center} 
\end{table}
The spontaneous breaking of R parity
and lepton number is driven by 
\be
v_R = \vev {\tilde{\nu}_{R\tau}} ~~~~~~~~~~~~
v_S = \vev {\tilde{S}_{\tau}} ~~~~~~~~~~~~
v_L = \vev {\tilde{\nu}_{\tau}}
\ee
The electroweak breaking and fermion masses arise from
\be
\vev {H_u} = v_u ~~~~~
\vev {H_d} = v_d
\ee
with $v^2 = v_u^2 + v_d^2$ fixed by the W mass.
The Majoron is then given by the imaginary part of 
\be
\frac{v_L^2}{V v^2} (v_u H_u - v_d H_d) +
              \frac{v_L}{V} \tilde{\nu_{\tau}} -
              \frac{v_R}{V} \tilde{{\nu^c}_{\tau}} +
              \frac{v_S}{V} \tilde{S_{\tau}}
\ee
where $V = \sqrt{v_R^2 + v_S^2}$. 
Since the majoron
is mainly an \21 singlet it does not contribute to the
invisible $Z^0$ decay width.

\subsection{Tree Level Breaking}

To study the breaking of R-parity in the model described in
Eq.~\ref{eq:super} we considered~\cite{pot}, for simplicity, 
the 1-generation case. The soft breaking terms are:
\begin{eqnarray}
V_{SB}
&=&
\tilde{m}_0 
\left[-A h_0 \Phi H_u H_d -  B 
\varepsilon^2 \Phi +C h_{\nu} \tilde{\nu}^c \tilde{\nu} H_u 
+  D h \Phi \tilde{\nu}^c S + h. c. \right]\nonumber \\ 
&&
+\tilde{m}_u^2 |H_u |^2 +\tilde{m}_d^2 |H_d |^2  
+\tilde{m}_L^2 |\tilde{\nu}|^2 +\tilde{m}_R^2 |\tilde{\nu}^c|^2
+\tilde{m}_S^2 |S|^2 + \tilde{m}_F^2 |\Phi |^2 
\end{eqnarray}
At unification scale we have~\footnote{In a N=1 SUGRA model.} $C=D=A 
\ ; \ B=A-2$ and universality of the soft masses 
$\tilde{m}_u^2=\tilde{m}_d^2= \cdots = \tilde{m}_0^2 $.
At low energy these relations will be modified by the 
renormalization group  evolution. For simplicity we take
\be
C=D=A \quad \hbox{and} \quad B=A-2
\ee
but let the soft masses at the weak scale be arbitrary.
Then the neutral scalar potential is given by
\begin{eqnarray}
V_{S}&=& {1 \over 8}\ (g^2+g'^2) \left[ |H_u |^2 - |H_d |^2 - 
|\tilde{\nu}|^2 \right]^2 + | h_0 \Phi H_u |^2
\nonumber \\
&& 
+| h \Phi S + h_{\nu} \tilde{\nu} H_u |^2
+|h \Phi \tilde{\nu}^c|^2 +| -h_0 \Phi H_d + h_{\nu} \tilde{\nu} 
\tilde{\nu}^c |^2 \nonumber \\ 
&&
+ |h \Phi \tilde{\nu}^c|^2 
+ |-h_0 H_u H_d + h \tilde{\nu}^c S -\varepsilon^2 
|^2 +V_{SB}
\end{eqnarray}
To find the solutions of the stationary equations we follow the 
following {\it 3 step } procedure~\cite{pot}:

\begin{enumerate}

\item
{\sl Finding solutions of the extremum equations}

We start by taking random values for 
$h$, $h_0$, $h_{\nu}$, $A$, $\varepsilon^2$ and $\tilde{m}_0$ and
$v_R$, $v_S$.
Then choose $\tan \beta={v_u \over v_d}$ and fix 
$v_u,v_d$ by $W$ mass relation,
\be
m_W^2=\frac{1}{2} g^2 (v_u^2+v_d^2+v_L^2)
\ee
Finally we solve the extremum equations {\sl exactly} for 
$\tilde{m}_u^2$, $\tilde{m}_d^2$, $\ldots$, $\tilde{m}_0^2 $. This is 
possible because they are linear equations on the mass squared 
terms.

\item
{\sl  Showing that the solution is a minimum}

To show that the solution is a true minimum we 
calculate the squared mass matrices for the real and imaginary parts
of the scalar fields, $M^2_R$ and $M^2_I$ and find numerically the
eigenvalues. 
The solution is a minimum if all nonzero eigenvalues
are positive. A consistency check is that we should get two zero 
eigenvalues for $M_I^2$ corresponding to the Goldstone boson of 
the $Z^0$ and to the majoron $J$.

\item
{\sl Comparing with other minima}

There are three kinds of minima to which we should compare our 
solution. 
\bea
&\bullet& v_u=v_d=v_L=v_R=v_S=0 \quad ; \quad v_F\not=0 \nonumber \\
&\bullet&v_L=v_R=v_S=0 \quad ; \quad
v_u,v_d,v_F\not=0 \nonumber \\
&\bullet&v_u=v_d=v_L=0 \quad ; \quad
v_R,v_S,v_F\not=0 \nonumber 
\eea
As a final result we found a large region in 
parameter space where our solution that breaks $R_P$ and
$SU_{2} \otimes U(1)$ is an absolute minimum.

\end{enumerate}

\subsection{Radiative Breaking}

We considered~\cite{rprad} the theory characterized by the following
superpotential:

\bea
W&=& h_u u^c Q H_u + h_d d^c Q H_d + h_e e^c L H_d   \nonumber \\
&&+h_0 H_u H_d \Phi +
h_{\nu} \nu^c L H_u + 
h \Phi \nu^c S +
\lambda \Phi^3
\label{eq:rprad}
\eea
For this theory with the following boundary
conditions at unification,
\bea
&&A_u = A = A_0 = A_{\nu} = A_{\lambda} \:, \nonumber\\
&&M_{H_u}^2 = M_{H_d}^2 = M_{\nu_L}^2 = M_{u^c}^2 = M_{Q}^2 =m_0^2 \:,
\nonumber \\
&&M_{\nu^c}^2 = C_{\nu^c}  m_0^2 \ ;M_{S}^2 = C_{S}  m_0^2 \ ;
M_{\Phi}^2 = C_{\Phi}  m_0^2  \:,  \nonumber \\
&&M_3 = M_2 = M_1 = M_{1/2} 
\label{eq:unif}
\eea
we run the RGE from the unification scale 
$M_U \sim 10^{16}$ GeV down to the weak scale. In doing 
this we randomly give values at the unification scale.
After running the RGE we have a complete set of parameters, 
Yukawa couplings and soft-breaking masses $m^2_i(RGE)$ 
to study the minimization of the potential,
\be
V_{total}  = \sum_i \left| { \partial W \over \partial z_i} \right|^2
	+ V_D + V_{SB} + V_{RC}
\ee
To solve the extremum equations we use the method described before,
except that now the value of $v_u$ is determined from 
$m_{top}=h_t v_u$ for $m_{top}=175 \pm 5$ GeV and
$v_d$ and $\tan(\beta)$ are then determined by $m_W$.
After doing this we end up with a set of points for which:

\begin{enumerate}

\item
The Yukawa couplings and the gaugino mass terms are given by the RGE.

\item
For a given set of $m^2_i$ each point is also a solution of the minimization
of the potential that breaks R-Parity. 

\item
However,  the $m^2_i$ obtained by the minimization
of the potential differ from those obtained from the RGE $m^2_i(RGE)$. 

\end{enumerate}

\noindent
Our  goal is to find solutions that obey
\be
m^2_i=m^2_i(RGE) \quad \forall i
\ee
To do that we define a function
\be
\eta= max \left( \frac{m^2_i}{m^2_i(RGE)},\frac{m^2_i(RGE)}{m^2_i}
\right) \quad \forall i 
\ee
that has the property $\eta \ge 1$. We are then all set for a minimization 
procedure. We were not able to find solutions with strict
universality. But if we relaxed the universality conditions on the 
squared masses of the singlet fields, as shown in Eq.~\ref{eq:unif},
 we got plenty of solutions.

\section{Main Features of the Model}

\subsection{Chargino Mass Matrix}

The form of the chargino mass matrix~\cite{paulo} is common to 
a wide class of  SUSY models with spontaneously broken R-parity
and is given by

\be
\begin{array}{c|cccccccc}
& e^+_j & \tilde{H^+_u} & -i \tilde{W^+}\nonumber \cr
\hline
e_i & h_{e ij} v_d & - h_{\nu ij} v_{Rj} &  g v_{Li} \cr
\tilde{H^-_d} & - h_{e ij} v_{Li} & \mu & g v_d \cr
-i \tilde{W^-} & 0 &  g v_u & M_2 
\end{array}
\label{eq:gino}
\ee
As a consequence the usual charged leptons will mix with 
the MSSM charginos.

\subsection{Neutralino Mass Matrix}

Under reasonable approximations, we can truncate the neu\-tralino 
mass matrix so as to obtain an effective $7\times 7$ matrix~\cite{npb} 

\be
\begin{array}{c|cccccccc}
& {\nu}_i & \tilde{H}_u & \tilde{H}_d & -i \tilde{W}_3 & -i \tilde{B}
\nonumber \\
\hline
{\nu}_i & 0 & h_{\nu ij} v_{Rj} & 0 & \frac{\displaystyle g}{\displaystyle \sqrt{2}}
v_{Li} & -\frac{\displaystyle g'}{\displaystyle \sqrt{2}} v_{Li}\nonumber \\
\tilde{H}_u & h_{\nu ij} v_{Rj} & 0 & - \mu 
& -\frac{\displaystyle g}{\displaystyle \sqrt{2}} v_u & \frac{\displaystyle g'}{\displaystyle \sqrt{2}} v_u\nonumber \\
\tilde{H}_d & 0 & - \mu & 0 & \frac{\displaystyle g}{\displaystyle \sqrt{2}} v_d 
& -\frac{\displaystyle g'}{\displaystyle \sqrt{2}} v_d\nonumber \\
-i \tilde{W}_3 & \frac{\displaystyle g}{\displaystyle \sqrt{2}} v_{Li} 
& -\frac{\displaystyle g}{\displaystyle \sqrt{2}} v_u & \frac{\displaystyle g}{\displaystyle \sqrt{2}} v_d 
& M_2 & 0\nonumber \\
-i \tilde{B} & -\frac{\displaystyle g'}{\displaystyle \sqrt{2}} v_{Li} & \frac{\displaystyle g'}{\displaystyle
\sqrt{2}} v_u & -\frac{\displaystyle g'}{\displaystyle \sqrt{2}} v_d & 0 & M_1 \nonumber
\end{array}
\label{eq:nino}
\ee
This matrix induces mixing between the neutrinos, considered as
Majorana fermions, and the MSSM neutralinos. As a result of these
mixings, both the charged and neutral current couplings are modified
with respect to the MSSM. This will be very important in the
phenomenological applications.

\subsection{Experimental Constraints}

While studying the phenomenology of these models there are many
experimental constraints that have taken in
account~\cite{paulo,npb,concha}. These come from
many different types of experiments. LEP searches puts limits on 
chargino masses and also in the amount of new contributions both to
the total and invisible $Z^0$ width. From the hadron colliders
there are restrictions on the gluino mass.
Finally there are additional restrictions, which are 
more characteristic of broken R-parity models. They follow 
from laboratory experiments related to {\sl neutrino physics}, 
{\sl cosmology} and {\sl astrophysics}.
The most relevant are:
neutrino-less double beta decay, neutrino oscillation searches,
direct searches for anomalous peaks at $\pi$ and K meson decays,
the limit on the tau neutrino mass and cosmological limits on 
the $\nu_{\tau}$ lifetime and mass.

\section{Implications for Neutrino Physics}

Here we briefly summarize the main results for neutrino physics.

\begin{itemize}

\item
{\it Neutrinos have mass}

Neutrinos are massless at Lagrangian level but get mass from the
mixing with neutralinos.\cite{paulo,npb}

\item
{\it Neutrinos mix}

The coupling matrix $h_{\nu_{ij}}$ has to be non diagonal to allow
\be
\nu_{\tau} \ra \nu_{\mu} + J 
\ee
and therefore evading~\cite{npb} the {\it Critical Density Argument} against
$\nu's$ in the MeV range.

\item
{\it Avoiding BBN constraints on the $m_{\nu_{\tau}}$}

In the {\it SM} BBN arguments~\cite{bbnothers} rule out $\nu_{\tau}$ 
masses in the range
\be
0.5\ MeV < m_{\nu_{\tau}} < 35 MeV
\ee
We have shown~\cite{bbnpaper} that {\it SBRP} models can evade that 
constraint due to new annihilation channels
\be
\nu_{\tau} \nu_{\tau} \ra J J 
\ee

\end{itemize}

\section{R-Parity Violation at LEP I}

\subsection{Higgs Physics}

The structure of the neutral Higgs sector is more complicated then in
the {\it MSSM}. However the main points are simple.

\begin{itemize}

\item
{\it Reduced Production}

Like in the {\it MSSM} the coupling of the Higgs to the $Z^0$ is
reduced by a factor $\epsilon_B$
\be
\epsilon_B=\left\vert 
\frac{g_{ZZh}}{\vbox to 12pt {} g_{ZZh}^{SM}} \right\vert < 1
\ee

\item
{\it Invisible decay}

Unlike the SM and the MSSM where the Higgs decays mostly in 
$b\overline{b}$, here it can have {\it invisible} decay modes like
\be
H \ra J + J 
\ee
Depending on the parameters, the $BR(H\ra \hbox{invisible})$ can be large.
This will relax the mass limits obtained from LEP. 
We performed a model
independent analysis of the LEP data~\cite{higgs} 
taking $m_H$, $\varepsilon_B$
and $BR(H\ra \hbox{invisible})$ as independent parameters. 

\end{itemize}

\subsection{Chargino Production at the Z Peak}

The more important is the possibility of the decay 
\be
Z^0 \to \chi^{\pm} \tau^{\mp}
\ee
This decay is possible because $R_p$ is broken. We have 
shown~\cite{paulo,val95} that this branching ratio can be as high 
as $5\times 10^{-5}$. 
Another important point is that the chargino has different decay modes
with respect to the MSSM. 

\begin{itemize}
\item
3-body decay \hskip 1cm
$\chi \ra \chi^0 + f \overline{f'}$ 

\item
2-body decay \hskip 1cm $\chi \ra
\tau + J$
\end{itemize}

\subsection{Neutralino Production at the Z Peak}

We have developed an event generator that simulates the 
processes expected for the LEP collider at $\sqrt{s}=M_{Z}$. Its main
features are:

\begin{itemize}

\item
{\it Production}

As far as the production is concerned, our generator 
simulates the following processes at the $Z$ peak:

\bea
&&\ e^+ e^-\rightarrow \chi\nu \\
&&\ e^+ e^-\rightarrow \chi \chi
\eea

\item
{\it Decay}

The second step of the generation is the decay of the lightest
neutralino. The 2-body only contributes to the missing energy. The
3-body are:

\bea
&&\chi \rightarrow \nu_{\tau} Z^{*} \rightarrow \nu_{\tau}\ l^+l^- ,
\nu_{\tau} \nu \nu,\nu_{\tau} q_i \overline{q_i}\\
&&\chi \rightarrow \tau \ W^{*} \rightarrow  \tau \nu_i l_i,\tau q_u
\overline{q_d}
\eea

\item
{\it Hadronization}

The last step of our simulation is made calling the 
PYTHIA software for the final states with quarks.

\end{itemize}

\noindent
One of the cleanest and most interesting signals that can be 
studied is the process with missing transverse momentum + 
acoplanar muons pairs
\be
 p\!\!\!/_T +\mu^+ \mu^- 
\ee
The main source of 
background for this signal is the 
\be
Z \ra \mu^+\mu^- + \hbox{soft photons}
\ee
For definiteness we have imposed the cuts used by the OPAL 
experiment for their search for acoplanar dilepton events:
(a) We select events with two muons with at least for one of the muons
obeying $ |\cos\theta | $ less than 0.7. 
(b) The energy of each muon has to be greater than a $6\%$ of the beam
energy.
(c) The missing transverse momentum in the event must exceed
$6\%$ of the beam energy, $p\!\!\!/_T > 3 \:$ GeV.
(d) The acoplanarity angle (the angle between the projected
momenta of the two muons in the plane orthogonal to the 
beam direction) must exceed $20^o$. With these cuts we were able to
calculate the efficiencies of our processes.

\begin{figure}[ht]
\begin{tabular}{cc}
\mbox{\epsfig{file=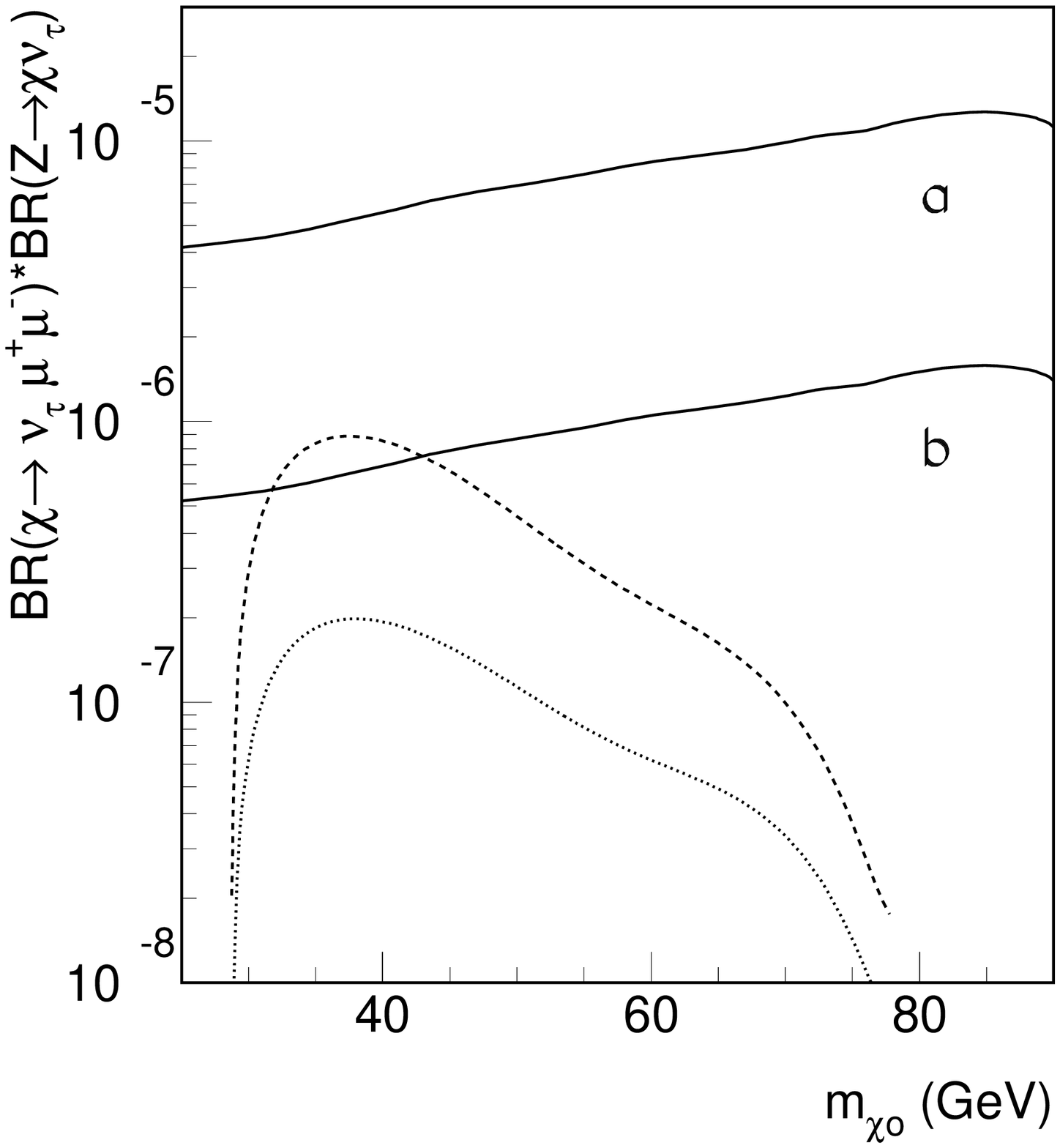,height=55mm,width=0.45\linewidth}}
&
\mbox{\epsfig{file=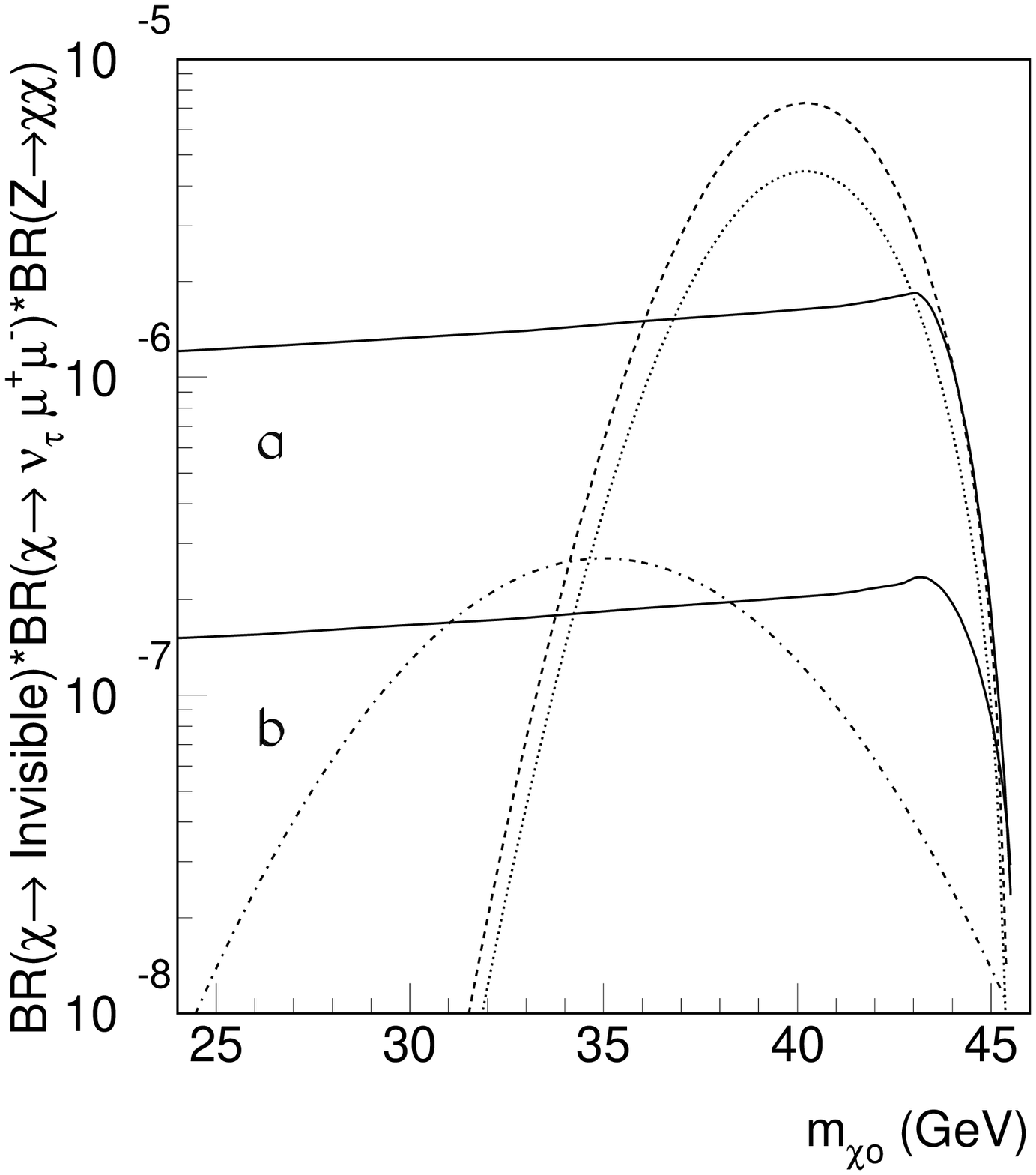,height=55mm,width=0.45\linewidth}}
\end{tabular}
\caption{On the left a comparison of the attainable limits on
$BR (Z \ra \chi \nu) BR(\chi \rightarrow \mu^+\mu^- \nu)$
versus the lightest neutralino mass, with the maximum theoretical values 
expected in different R-parity breaking models. The solid line (a) is
just for the $\mu^+\mu^-\nu$ channel, 
while (b) corresponds to the improvement expected
from including the $e^+e^-\nu$ channel, as well as the 
combined statistics of the four LEP experiments.
The dashed line corresponds to a model with explicit R-parity
violation,while the dotted one is calculated in the 
spontaneous R-parity-violation model. On the right the same for
$BR (Z \ra \chi \chi) BR(\chi \rightarrow \mu^+\mu^- \nu)$.
}
\label{fig1}
\end{figure}

\noindent
We used the data published by ALEPH in 95 and analyzed both the single
production $e^+e^- \rightarrow \chi \nu $ and the double production 
$e^+e^- \rightarrow \chi \chi$
processes. For single production we get
\be
N_{expt} (\chi \nu)  = \sigma(e^+e^-\rightarrow \chi \nu) BR(\chi \rightarrow 
\nu_\tau \mu^+\mu^-) \epsilon_{\chi \nu} \  L_{int}
\ee
Using the expression for the cross section we can write this
expression in terms of the product $BR (Z \ra \chi \nu)$ $\times$
$BR(\chi \rightarrow \nu_\tau \mu^+ \mu^-)$
and obtain a 
$95\% CL $ limit on this R-parity breaking observable,
as a function of the $\chi$ mass. This is shown in 
Figure~\ref{fig1}. For the double production of neutralinos the number
of expected $p\!\!\!/_T+\mu^+ \mu^-$  events is
\be
N_{expt} (\chi \chi) = \sigma (e^+e^- \rightarrow \chi \chi)
2 BR (\chi \rightarrow \mbox{invisible})
BR(\chi\rightarrow\nu_\tau\mu^+\mu^-) \epsilon_{\chi \chi} \  L_{int}
\ee
We can obtain 
an illustrative $95\% CL$ limit on $BR (Z \ra \chi \chi)$ $\times$
$BR (\chi \ra \nu_\tau \mu^+\mu^-)$ $\times$ $ BR (\chi \ra \mbox{invisible})$ 
as a function of the $\chi$ mass. This is also shown in Figure \ref{fig1}
where we can see that the models begin to be constrained by the LEP results.

\section{R-Parity Violation at LEP II}

\subsection{Invisible Higgs}

The previous LEP I analysis has been extended for LEP II.\cite{eboli}
As a general framework we consider models with the  interactions

\begin{eqnarray}
{\cal L}_{hZZ}
&=& \epsilon_B \left ( \sqrt{2}~ G_F \right )^{1/2}
M_Z^2 Z_\mu Z^\mu h 
\; , 
\nonumber \\
{\cal L}_{hAZ}&=& - \epsilon_A \frac{g}{\cos\theta_W} 
Z^\mu h \stackrel{\leftrightarrow}{\partial_\mu} A 
\; ,
\end{eqnarray}
with $\epsilon_{A(B)}$ being determined once a model is chosen. We
also consider the possibility that the Higgs decays invisible
\be
h \ra JJ
\ee
and treat the branching fraction $B$ for $h \rightarrow JJ$ as a 
free parameter.

\begin{figure}[ht]
\begin{tabular}{p{0.45\linewidth}p{0.45\linewidth}}
\mbox{\epsfig{file=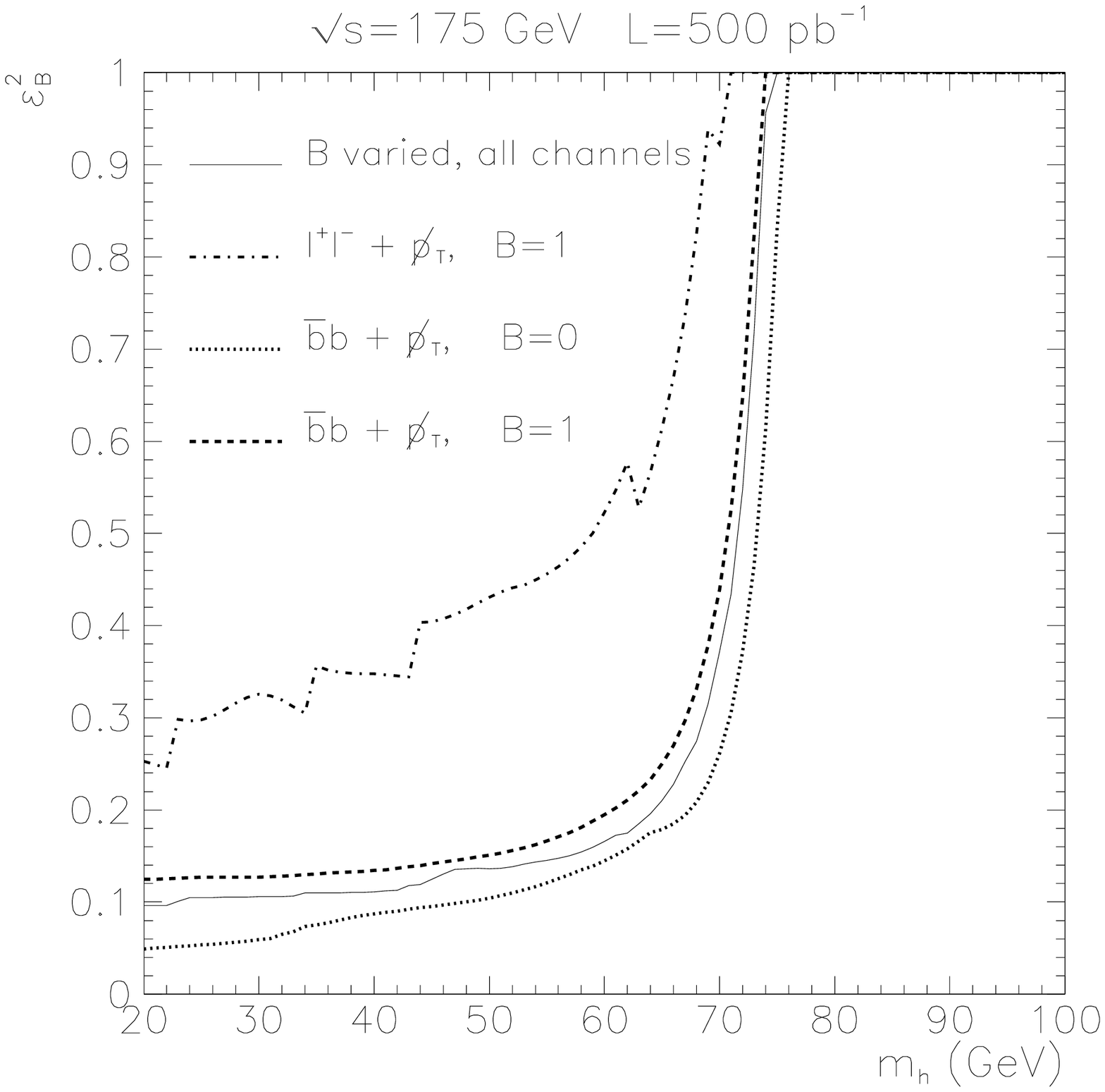,width=0.95\linewidth,bbllx=10pt,bblly=10pt,%
bburx=530pt,bbury=530pt}}
&
\mbox{\epsfig{file=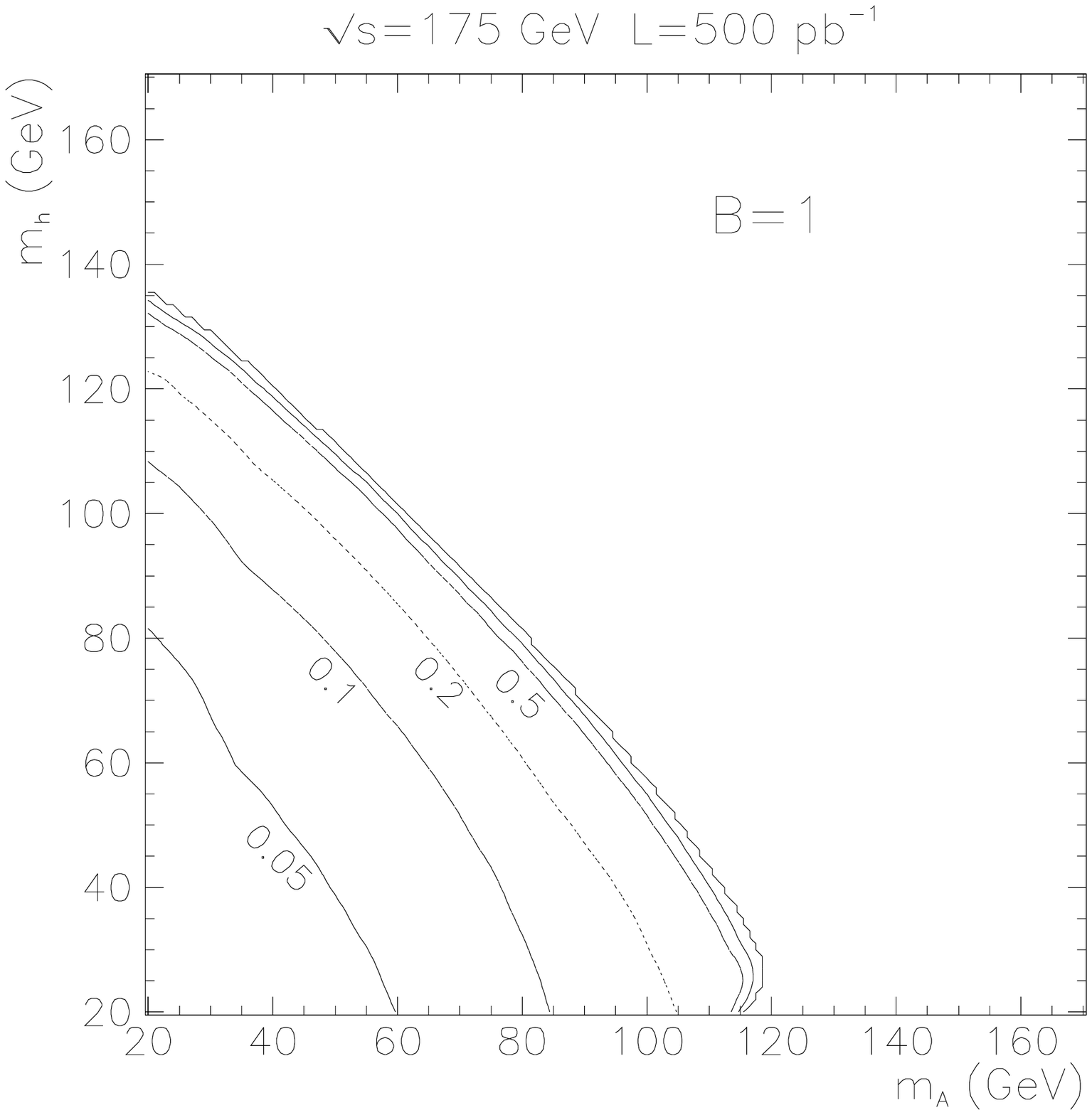,width=0.95\linewidth,bbllx=10pt,bblly=10pt,%
bburx=530pt,bbury=530pt}}
\end{tabular}
\caption{On the left, bounds on $\epsilon_B^2$ as a function of $M_h$
for $\sqrt{s}=175 GeV$. On the right, bounds on $\epsilon_A^2$ as a
function of $M_h$ and $ M_A$ for $B=1$ and $\surd{s}= 175$ GeV.}
\label{fig3}
\end{figure}

\noindent
The following signals with $\ptmis$ were considered:

\begin{eqnarray}
e^+ e^-  &\rightarrow & (Z h + A h) \rightarrow b \bar{b}~+~ \ptmis
\; ,
\nonumber \\
e^+ e^-  &\rightarrow & Z h \rightarrow \ell^+ \ell^- ~+~ \ptmis
\; ,
\end{eqnarray}
but also the more standard processes
\begin{eqnarray}
e^+ e^-  &\rightarrow & Z h \rightarrow \ell^+ \ell^-  +~b \bar{b} 
\; ,
\nonumber \\
e^+ e^-  &\rightarrow & (Z h + A h) \rightarrow b \bar{b}~ +~b \bar{b}
\; .
\end{eqnarray}
Using the above processes and after a careful study of the backgrounds
and of the necessary cuts,~\cite{eboli} it was possible to 
evaluate the limits on $M_h$, $M_A$, $\epsilon_A$,
$\epsilon_B$, and $B$ that can be obtained at LEP II. In
Figure~\ref{fig3}  are shown some of these limits.

\ignore{
\begin{figure}[ht]
\begin{tabular}{p{0.45\linewidth}p{0.45\linewidth}}
\mbox{\epsfig{file=iah_175.ps,width=0.95\linewidth,bbllx=10pt,bblly=10pt,%
bburx=530pt,bbury=530pt}}
&
\mbox{\epsfig{file=ah_175.ps,width=0.95\linewidth,bbllx=10pt,bblly=10pt,%
bburx=530pt,bbury=530pt}}
\end{tabular}
\caption{Bounds on $\epsilon_A^2$ as a function of $M_h$ and $ M_A$. The plot
on the left is for for $B=1$ and $\surd{s}= 175$ GeV. The plot on the
right shows $B$-independent bounds on $\epsilon_A^2$  for the same
conditions. The allowed region of the parameter space is above the lines
of constant $\epsilon_A$.}
\label{fig4}
\end{figure}
}

\subsection{Neutralinos and Charginos}

At LEP II the production rates for {\it R-Parity violation} processes
will not be very large, compared with
those at LEP I. Therefore we expect that the production rates will be like
in the MSSM, via non R-parity breaking processes. 
However the decays will be
modified much in the same way as in the LEP I case. This is specially
important for the $\chi_0$ because it is invisible in the MSSM but visible
here. 
Also the R-parity violating decays of the charginos
\be
\chi^- \ra \tau^- + J
\ee
can have a substantial decay fraction compared with the usual MSSM decays
\be
\chi^- \ra \chi^0 + f \overline{f'}
\ee

\section{Conclusions}

There is a viable model for {\bf SBRP} that leads to a very rich
phenomenology, both at laboratory experiments, and at present (LEP)
and future (LHC, LNC) accelerators.
We have shown that the radiative breaking of {\bf both} the Gauge
Symmetry and R-Parity can be achieved.
In these type of models neutrinos have mass and can decay thus
avoiding the critical density argument. They also can evade the BBN
limits on a $\nu_{\tau}$ on the MeV scale.
Regarding Higgs Physics, the most important point is that the lightest
Higgs boson can have a significant invisible decay fraction. This
changes the standard analysis for the Higgs mass limits.
We have illustrated how the existing 
data gathered by the LEP collaborations at the Z peak are sufficient 
to probe the spontaneously broken R-parity models. These results have
been extended for the case of LEP II after they have
collected ${\cal L}= 500 pb^{-1}$ of data.

\section*{Acknowledgments}
This work was supported in part by the TMR network 
grant ERBFMRX-CT960090 of the European Union.

\end{document}